# MODEL TO STUDY INTERPLAY BETWEEN BEAMSTRAHLUNG AND COUPLING IMPEDANCE IN FUTURE LEPTON COLLIDERS

D. Leshenok[1] and S. Nikitin[2]
*Budker Institute of Nuclear Physics, 630090 Novosibirsk, Russia*
M. Zobov[3]
*INFN, Laboratori Nazionali di Frascati, via Enrico Fermi 40, 00044 Frascati, RM, Italy*

*Abstract*

A semi-analytical model has been developed to study a combined effect of Beamstrahlung due to beam-beam interaction and beam coupling impedance in the future lepton colliders CEPC and FCCee. This model allows evaluating an impact of the coupling impedance on the bunch length and energy spread in collision. The model is benchmarked against numerical simulations. Analytical estimates for the supercolliders are presented.

## INTRODUCTION

In order to perform very high precision physics experiments the future supercolliders FCCee [1] and CEPC [2] are going to exploit the Crab Waist (CW) collision scheme [3, 4] to push their luminosity. Due to tiny sizes of intense beams to be used in collision the Beamstrahlung (BS) effect becomes important leading to a substantial bunch lengthening and energy spread increase [5]. In the crab waist collisions with a large Piwinski angle the bunch length increase results in the luminosity reduction while the energy spread affects the experiment energy resolution. In turn, the beam coupling impedance is also responsible of the bunch lengthening and in addition it can result in the energy spread increase if the microwave instability threshold is exceeded. Moreover, the imaginary part of the impedance reduces the synchrotron frequency due to the potential well effect (see, for example [6]) and the collision working point is to be re-optimized to compensate this shift.

The interplay of the both effects can change substantially final values of the bunch length, the energy spread and the synchrotron tune. For example, the energy spread due to BS can suppress the microwave instability and the longer bunch will produce a weaker wake potential. On other hand, the impedance related bunch lengthening is expected to reduce the BS effect strength.

The 3D self-consistent numerical simulations of the combined effect of beam-beam interaction and the beam coupling impedance require a huge computational time [7]. In this paper we present a semi-analytical model that allows evaluating the bunch length, energy spread and the synchrotron frequency in a much shorter time. The results obtained for FCCee and CEPC by using the model are compared with available numerical simulations data.

___________________
[1] dariya210612@gmail.com; [2] nikitins@inp.nsk.su;
[3] Mikhail.Zobov@inf.infn.it

## ENERGY LOSS DUE TO BS

Let the electron and positron bunches colliding at the crossing angle θ << 1 contain N particles each, and σ$_x$, σ$_y$, and σ$_z$ are their horizontal, vertical, and longitudinal sizes, respectively (see Fig.1)

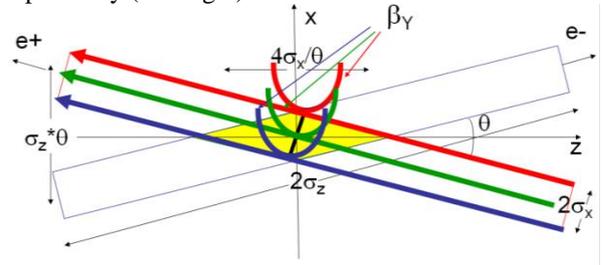

Figure 2 — Collision with a large Piwinski angle

In order to estimate the average particle energy loss due to radiation in the collective field of the oncoming bunch (BS), we use the approach developed by one of the author in [8, 9]. It is based on using the Takayama's potential [10] of a 3D Gaussian charge distribution. The BS energy losses are calculated in the rest frame of the oncoming bunch, and then recounted in the laboratory system. For the isolated case of a head-on collision, loss averaging is performed over all particles of the test bunch. For a nonzero but small crossing angle, averaging is limited, for simplification, by the region of the central cross section of the bunch. In the latter case, the ratio of the average loss to the initial particle energy is found from the equation (σ$_z$ >> σ$_x$, σ$_y$) [9].

$$U_{BS} = \langle \frac{\Delta E_{BS}}{E}\rangle \approx \frac{4}{3}\frac{r_e^3 N^2 \gamma}{\sqrt{\pi}\sigma_z} \iint_0^\infty \frac{dt dt'}{\Omega\sqrt{(2\sigma_x^2+t)(2\sigma_x^2+t')}} \cdot \left[\frac{\sigma_x^2}{g_x\sqrt{g_y}}\left(1+\frac{\theta^2 \sigma_z^2}{8\sigma_x^2 \Omega^2}\right) + \frac{\sigma_y^2}{g_y^{3/2}}\right], \quad (1)$$

Here $r_e$ is the classical electron radius; γ - relativistic factor of beam particles;
$g_x(t,t') = (2\sigma_x^2+t)(2\sigma_x^2+t') + 2\sigma_x^2(4\sigma_x^2+t+t')$,
$g_y(t,t') = (2\sigma_y^2+t)(2\sigma_y^2+t') + 2\sigma_y^2(4\sigma_y^2+t+t')$;
the functional

$$\Omega = \sqrt{1+\left[\frac{1}{2\sigma_x^2+t}+\frac{1}{2\sigma_x^2+t'}\right]\frac{\sigma_z^2 \theta^2}{4}} \quad (2)$$

represents a generalized Piwinski factor. When $t = t' = 0$, Ω takes the known form:

$$\Omega(0,0) = \sqrt{1 + \left(\frac{\sigma_z \theta}{2\sigma_x}\right)^2} \approx \sqrt{1 + \Psi^2}.$$

The numerical evaluation of the double integral in (1) takes less than a minute on a conventional PC. The formula (1) allows estimating the influence of BS on the formation of the longitudinal beam size and energy spread in supercolliders [11]. To this aim, the approximation of the interaction region in the form of an equivalent magnet with some effective values of the field and length can serve as the simplest model. In the theory of synchrotron radiation, losses in such a magnet are proportional to the product of the square of its field by the length. At large Piwinski "angle" $\Psi \gg 1$ as it takes a place in the Crab-Waist interaction region of the supercolliders, the size $L_{eff} = 4\sigma_x/\theta$ (see Fig.1)) can be approximately considered as a full effective length of the "magnet".

The model assumes that the beams retain their shape and sizes. In reality, there is an increase of the transverse bunch sizes at distances comparable with or larger than $\beta_y^*$ ($\beta_x^*$), the vertical (horizontal) beta function value at IP (hourglass effect). This occurs vertically, starting from rather shorter distances, in comparison with the horizontal direction, since the vertical beta function is much smaller than the horizontal one. To verify this effect, the growth rate of losses was calculated depending on the distance of the center of the test bunch to the IP with and without taking into account the corresponding change in vertical size at the FCCee example [9]. This numerical experiment demonstrated the almost complete insensitivity of the model to the hour-glass, which can be explained by the hierarchy of bunch sizes and the fact that the beams intersect in the median plane.

## BUNCH ENERGY SPREAD AND LENGTH WITH BS

To determine the steady-state beam energy spread and **a** self-consistent estimate of the length $\sigma_z$ as a result of BS, we use the equation of radiative diffusion of the particle energy taking into account radiative damping [12]:

$$\frac{d\langle A^2\rangle}{dt} = D_\gamma - 2\frac{\langle A^2\rangle}{\tau_E}. \quad (3)$$

Here $\langle A^2\rangle$ is the average square of the energy oscillation amplitude in a beam; $\tau_E$ is the corresponding damping time; $D_\gamma$ is a sum of the coefficients of diffusion due to synchrotron radiation (SR) in the bending magnets and BS in the interaction points (IP): $D_\gamma = D_{BS} + D_{SR}$. The $D_{SR}$ value was taken based on the design data of the collider magnetic structure. The $D_{BS}$ was estimated by the formula:

$$D_{BS} = n_{IP} f_0 C_u u_c P_\gamma \frac{L_{eff}}{c} \quad (4)$$

with $n_{IP}$ a number of the interaction points which are considered identical to each other (in our case $n_{IP} = 2$); $f_0$ the revolution frequency; $C_u = 1.32$; $u_c = 3\hbar c\gamma^3/(2\rho)$ the characteristic radiation quantum in the "long magnet" approximation; $P_\gamma = cEU_{BS}/L_{eff}$ the power defined through the energy loss for BS (4); $L_{eff}$ the "equivalent magnet" length determined in the section above; $\rho = \sqrt{2\pi P_\gamma/(cC_\gamma E^4)}$ the curvature radius of particle trajectory in the "equivalent magnet". In study of the BS spectrum, it was shown [13] that the approximation of a "long" (ordinary) magnet still remains relevant as applied to the present projects of supercollider. It does not need to be changed to the "short magnet" model.

Since $U_{BS}$ in (1) depends on the beam length $\sigma_z$, and that in turn is proportional to the energy spread $\sigma_E$, the stationary solution (3) is found from the transcendental equation

$$\sigma_E^2 = \frac{1}{4E^2}\tau_E D_\gamma(\sigma_z), \quad (5)$$

with

$$\sigma_z = \frac{c\alpha_p}{\nu_s \omega_0} \cdot \sigma_E, \quad (6)$$

$\alpha_p$ is the momentum compaction factor and $\nu_s$ is the synchrotron tune, $\omega_0 = 2\pi f_0$. The system of equations (4) - (6) is solved iteratively, which gives self-consistent values of the beam length and energy spread taking BS into account. It is assumed that the beam transverse dimensions at IP remain constant. This is true, firstly, because with the nominal beam parameters the beam-beam limit is not reached. Secondly, due to zero dispersion in the interaction region, BS does not lead to an additional increase in the transverse emittances.

The results of our calculations of the energy spread and the bunch length at the energies 45, 80 and 120 GeV for CEPC and FCC-ee are summarized in Table 1 and Table 2, respectively. For comparison in Table 1 we give the data provided by Yuan Zhang for CEPC [14] while in Table 2 the results for FCCee calculated by Dmitry Shatilov using the known LIFETRAC code [15,16] are shown.

Table 1.

| | CEPC | | | |
|---|---|---|---|---|
| E[GeV] | 45.5 Z(2T) | 45.5 Z(3T) | 80 W | 120 Higgs |
| Beam-Beam simulation by Y. Zhang | | | | |
| $\sigma_E$ | 0.00110 | | --- | 0.00146 |
| $\sigma_z$, mm | 7.0 | | --- | 4.0 |
| Semi-analytical model (SR+BS) | | | | |
| $\sigma_E$ | 0.00108 | 0.00108 | 0.00111 | 0.00137 |
| $\sigma_z$ mm | 6.8 | 6.8 | 4.9 | 3.7 |

Table 2.

| | FCCee | | |
|---|---|---|---|
| E [GeV] | 45.6 | 80 | 120 |
| FCCee CDR | | | |
| $\sigma_E$ | 0.00132 | 0.00131 | 0.00165 |
| $\sigma_z$ [mm] | 12.1 | 6.0 | 5.3 |
| Semi-analytical model (SR+BS) | | | |
| $\sigma_E$ | 0.00135 | 0.00130 | 0.00166 |
| $\sigma_z$ [mm] | 12.4 | 5.9 | 5.3 |

The reported numerical and analytical results are in a satisfactory agreement. However, these values do not take into account the bunch lengthening due to the beam coupling impedance.

## COMBINED MODEL WITH BUNCH LENGTHENING

Since most of the bunch lengthening in both CEPC [17] and FCCee [18] is due to the potential well distortion below the microwave instability threshold, at each iteration step we can include the impedance related bunch lengthening by calculating the synchrotron frequency variation in (6) due to the imaginary part of the impedance [19]

$$\nu_s^2 = \nu_{s0}^2 \left(1 - \frac{\xi}{2\pi} Z_{pot}\right), \quad (7)$$

where [20]

$$Z_{pot} = \sum_{p=-\infty}^{\infty} \mathrm{Im} Z(p\omega_0) \frac{J_1(p\omega_0 \hat\tau)}{\omega_0 \hat\tau / 2} \hat\lambda(p\omega_0),$$

with $\hat\lambda$ the Fourier transform of bunch line density and $\hat\tau$ the amplitude of synchrotron oscillations.

If for simplicity we assume that the total bunch lengthening is determined by pure inductive impedance then the following self-consistent solution can be found. For a Gaussian bunch

$$Z_{pot} = \frac{\sqrt{2\pi}}{(\omega_0 \sigma)^3} \cdot Z_c, \quad (8)$$

$\sigma = \sigma_z/c$. The quantity $Z_c$ does not depend on bunch lengthening, i.e. remains constant, and can be determined from the equation describing the lengthening effect in the case of no beam collision [19]:

$$x_0^3 - x_0 - \frac{\xi}{2\pi} Z_c \cdot \frac{\sqrt{2\pi}}{(\omega_0 \sigma_{SR})^3} = 0 \quad (9)$$

with $x_0 = \sigma_0/\sigma_{SR}$, any known ratio of the increased beam length $\sigma_{z0} = c\sigma_0$ to the length $\sigma_{z,SR} = c\sigma_{SR}$ determined by radiation in bending magnets (SR). The parameter $\xi$ depends on the bunch current $I_b$:

$$\xi = \alpha_p e I_b / \nu_s^2 E. \quad (10)$$

So, the $Z_c$ constant is defined by the equation

$$Z_c = \frac{\sqrt{2\pi}}{\xi}(x_0^3 - x_0)(\omega_0 \sigma_{SR})^3. \quad (11)$$

Below the microwave threshold the inductive impedance decreases the synchrotron tune and, accordingly, increases the bunch length without affecting the energy spread. At the same time, the BS seeks to increase the energy spread and thereby lengthens the bunch, depending on the synchrotron tune variation. Under equilibrium conditions, there must be a self-consistent solution for all the three parameters of that interplay. Let $x = \sigma/\sigma_{SR}$ be a steady-state lengthening ratio in such conditions. Considering (8) and (11), the equation (7) can be rewritten in the form

$$\nu_s^2 = \nu_{s0}^2 \left(1 - \frac{x_0^3 - x_0}{x^3}\right). \quad (12)$$

Given the contribution of BS to the diffusion coefficient, we have from the equation (6)

$$x^2 = \frac{D_{BS} + D_{SR}}{D_{SR}} \cdot \frac{\nu_{s0}^2}{\nu_s^2} = \eta \frac{\nu_{s0}^2}{\nu_s^2}. \quad (13)$$

Combining (12) and (13) gives an equation for a self-consistent solution for lengthening a bunch:

$$\frac{x^3 - \eta x}{x_0^3 - x_0} = 1. \quad (14)$$

It is easy to see that at $\eta = 1$ (no BS) the equation (14) takes a well-known form, describing the lengthening versus bunch current below the turbulence threshold ($x_0^3 - x_0 \propto I_b$).

In principle, an iterative algorithm to solve the problem in the described approach includes the following steps:
1. At the first step, set $x = x_0$ using the available data on bunch lengthening in CEPC and FCCee obtained at the nominal bunch current as a result of only PWD, i.e. without taking into account BS;
2. From (1), calculate the BS loss with a current value of the bunch length $\sigma_z^{(i)} = x^{(i)} \sigma_{z,SR}$;
3. Using (3)-(5), find the corresponding energy spread $\sigma_E^{(i)}$ in the presence of BS;
4. Re-calculate the synchrotron tune from (12);
5. Calculate a new value of the bunch length $\sigma_z^{(i+1)}$ from (6);
6. Based on a comparison of $\sigma_z^{(i)}$ and $\sigma_z^{(i+1)}$, either repeat the cycle from the point 2 or end the iteration.

A step-by-step search of a solution takes quite a long time. Therefore, by setting the initial length in the suspected region of existence of the solution, the calculation time can be reduced to several minutes.

The described algorithm corresponds to a situation when the beams at the nominal currents are brought into collision. An alternative is the sequential accumulation of currents to nominal values in colliding beams (the bootstrapping mode). The algorithm will be complicated by the need to recalculate the magnitude of the impedance at each step of the current increment.

Our calculations taking into account the combined effect of both BS and PWD effects were performed for CEPC at energies of 45.5, 80, and 120 GeV (see Table 3) with the data on lengthening due to PWD [2]. For instance, at 45.5 GeV, the latter is from 2.42 to 5.1 mm. For FCCee, due to the lack of similar data for 80 and 120 GeV, the calculations were carried out only for 45.6 GeV (see Table 4). The lengthening due to PWD is from 3.5 to 6 mm [6]. For comparison, both tables (Table 3 and Table 4) show also the results of the simulation by Y. Zhang [14] accounting beam lengthening due to the coupling impedance. In the tables, the beam length and energy spread are given taking into account the already complete energy losses on SR and BS. As can be seen from the comparison, even despite such a simplification used, the results of semi-analytical calculations are in reasonable agreement with numerical modeling.

## CONCLUSIONS

Using an approximate method that does not require the use of dedicated beam-beam simulation codes, a self-consistent stationary solution of the energy diffusion equation with BS is found.

In addition, an algorithm has been developed for calculating the Beamstrahlung effect taking into account bunch lengthening caused by the potential well distortion due to beam coupling impedance.

Table 3.

| CEPC | | | | |
|---|---|---|---|---|
| E GeV | 45.5 Z(2T) | 45.5 Z(3T) | 80 | 120 |
| Beam-beam simulation with coupling impedance [14] | | | | |
| $\sigma_E$ | --- | --- | --- | 0.00137 |
| $\sigma_z$ mm | --- | --- | --- | 4.1 |
| Semi-analytical model (SR+BS+PWD) | | | | |
| $v_s/v_{s0}$ | 0.859 | 0.858 | 0.868 | 0.904 |
| $\sigma_E$ | 0.000996 | 0.000993 | 0.00104 | 0.00132 |
| $\sigma_z$ mm | 7.3 | 7.3 | 5.3 | 4.0 |

Table 4.

| FCCee | |
|---|---|
| Energy [GeV] | 45.6 |
| Beam-beam simulation with coupling impedance [14] | |
| $\sigma_E$ | 0.00126 |
| $\sigma_z$ [mm] | 12.2 |
| Semi-analytical model (SR+BS+PWD) | |
| $v_s/v_{s0}$ | 0.964 |
| $\sigma_E$ | 0.00132 |
| $\sigma_z$ [mm] | 12.6 |

The proposed semi-analytical model has been used to calculate the equilibrium values of the bunch length and energy spread as well as the synchrotron frequency reduction for the CEPC and FCCee colliders at their various target energies.

A comparison of the data calculated by using the model with the available data obtained previously with the beam-beam simulation codes has been made. The differences in the beam length (and energy spread), taking into account only BS, are 1.5% for CEPC and about 2% for FCCee. Given the combined effects of BS and PWD effects, these differences do not exceed 3% for CEPC and 3.5% for FCCee, respectively.

Thus, the proposed method allows estimating accurately the BS effect, alone and in combination with PWD, on the longitudinal beam size, energy spread and synchrotron frequency reduction without resorting to a full-scale beam-beam simulation.

## ACKNOWLEDGMENTS

Authors thanks Dmitry Shatilov for useful discussions and Y. Zhang for the information on his recent results in beam-beam simulation.

## REFERENCES

[1] A. Abada et al., "FCC-ee: the lepton collider: future circular collider conceptual design report volume 2", *Eur. Phys. J. Spec. Top.*, vol. 228, pp. 261–623, Jun. 2019.
doi:10.1140/epjst/e2019-900045-4

[2] CEPC Study Group, "CEPC conceptual design report: volume 1 – accelerator", e-Print: arXiv: 1809.00285, Sep.2018.

[3] P. Raimondi, D. Shatilov, and M. Zobov, "Beam-beam issues for colliding schemes with large Piwinski angle and crabbed waist", Rep. LNF-07-003-IR, e-Print: physics/0702033, Feb. 2007.

[4] M. Zobov et al., "Test of crab-waist collisions at DAΦNE Phi factory", *Phys. Rev. Lett.*, vol. 104, p. 174801, Apr. 2010.
doi: 10.1103/PhysRevLett.104.174801

[5] V. Telnov, "Restriction on the energy and luminosity of e+e- storage rings due to beamstrahlung", *Phys. Rev. Lett.* vol. 110, p. 114801, Mar. 2013.
doi: 10.1103/PhysRevLett.110.114801

[6] E. Belli, M. Migliorati, and M. Zobov, "Impact of the resistive wall impedance on beam dynamics in the Future Circular e+e− Collider", *Phys. Rev. Accel. Beams,* vol. 21, p. 041001, Apr. 2018.
doi: 10.1103/PhysRevAccelBeams.21.041001

[7] Y. Zhang, C. T. Lin, N. Wang, and C. H. Yu, "Crosstalk of Beam-Beam Effect and Longitudinal Impedance at CEPC", in *Proc. 10th Int. Particle Accelerator Conf. (IPAC'19)*, Melbourne, Australia, May 2019, pp. 247-249
doi:10.18429/JACoW-IPAC2019-MOPGW068

[8] E. Levichev, and S. Nikitin, "Concept of waveguide Compton monitor of beam energy in high energy e+e- collider", *JINST,* vol. 11, p. P06005, Jun. 2016
doi: 10.1088/1748-0221/11/06/P06005

[9] S. Nikitin, "Formulae for estimating average particle energy loss due to Beamstrahlung in supercolliders", e-Print: arXiv: 2002.01173, Feb.2020.

[10] K. Takayama, "Potential of a 3-dimensional halo charge distribution", *IEEE Trans. Nucl. Sci.,* vol. 30, p. 2661, Mar. 1983.  doi: 10.1109/TNS.1983.4332916

[11] D. Leshenok, and S. Nikitin, "Effect of energy loss on accuracy of precision energy measurements in CEPC", presented at the CEPC Workshop, Beijing, China, Nov. 2019, unpublished.

[12] M.Sands, "The physics of electron storage rings: an introduction", Rep. SLAC-R-121, Nov. 1970.

[13] V. Telnov, Talk at IPAC2018, Vancouver, 2018,
https://accelconf.web.cern.ch/accelconf/ipac2018/talks/weygbe3_talk.pdf

[14] Y. Zhang, private communication.

[15] D. Shatilov, "Beam-beam simulations at large amplitudes and lifetime determination", Part. Accel., vol.52, pp. 65-93, Jul. 1996.

[16] D. Shatilov, "FCC-ee parameter optimization", *ICFA Beam Dyn. Newslett.*, vol. 72, pp. 30-41, 2017.

[17] Na Wang et al., "Mitigation of coherent beam instabilities in CEPC", presented at ICFA mini-Workshop on MBCI, Zermatt, Switzerland, Sep. 2019.

[18] E. Belli et al., "Electron cloud buildup and impedance effects on beam dynamics in the future circular e+e- collider and experimental characterization of thin TiZrV vacuum chamber coatings", *Phys. Rev. Accel. Beams,* vol. 21, p. 011002, Nov. 2018.

[19] Handbook of accelerator physics and engineering, edited. by A. W. Chao, K.H. Mess, M. Tigner, and F. Zimmermann, World Scientific (2013).

[20] Handbook of accelerator physics and engineering/edited by A.W. Chao and M. Tigner, 650 pages, World Scientific (1998).